\documentclass[twocolumn,showpacs,preprintnumbers,amsmath,amssymb,prb]{revtex4-1}
\usepackage{graphicx}  
\usepackage{dcolumn}   
\usepackage{bm}        
\usepackage[utf8]{inputenc}
\usepackage{graphics}
\usepackage{latexsym}
\usepackage{amsmath}
\usepackage{float}
\usepackage{cancel}
\usepackage{multirow} 
\usepackage{pstricks,pst-plot}
\newcommand{\bra}[1]{\ensuremath{\langle #1 |}}
\newcommand{\ket}[1]{\ensuremath{| #1 \rangle}}

\begin{document}


\title{Transverse current response of graphene at finite temperature, plasmons and absorption}
\author{A. Guti\'errez-Rubio${}^1$, T. Stauber${}^2$, and F. Guinea${}^1$}
\affiliation{${}^1$ICMM - CSIC, E-28049 Madrid, Spain\\
${}^2$Departamento de F\'{\i}sica de la Materia Condensada  and Instituto Nicol\'as Cabrera, Universidad Aut\'onoma de Madrid, E-28049 Madrid, Spain}
\date{\today}

\begin{abstract}
We calculate the linear transverse current current response function for graphene at finite temperature and chemical potential. Within the Random Phase Approximation, we then discuss general aspects of transverse plasmons beyond the local response such as their dependence on temperature and on the surrounding dielectric media. We find, e.g., maximal confinement of this mode for a homogeneous dielectric media with refractive index $n\simeq 40$. Confinement can further be enhanced by placing the graphene sheet inside an optical cavity, but there exists a critical width below which no transverse mode can be sustained. For zero doping and finite temperature, there are no well-defined transverse plasmonic excitations in contrast to the longitudinal channel. We also discuss the absorption of electromagnetic radiation in single and double-layer systems for $s$ and $p$ polarizations and point out that the theoretical limit of 50\% is reached for $s$-polarized light with incident angle of $\theta\approx89^o$.
\end{abstract}

\maketitle


\section{Introduction}

During the last years, plasmonics has become a celebrated field to which deep and intense research is being targeted. The term plasmon alludes to a collective harmonic movement of electrons\cite{Ritchie} entailed to equally oscillating electromagnetic fields, and was born in the context of studying losses of fast electrons when moving through metallic foils. \cite{Pines,Bohm52} This concept also emerges as the existence of electromagnetic modes at the interface between a metal and a dielectric.\cite{Otto68,Kretschmann71} 

Its popularity lies in both the theoretical and practical interest that its understanding offers. On the one hand, they manifest many-body effects between electrons associated to the long range Coulomb interaction;\cite{Maier07,GarciaVidal10} on the other hand, their striking features, among which their high confinement to the two-dimensional sheet plays a main role, suggest various applications in cutting-edge electronics and optoelectronics.\cite{Ozbay06} This altogether enforces the will to extend this exploration to systems of present physical interest that share a challenge in the theoretical aspect with a promise regarding engineering or applied physics.

Graphene is one of the most promising candidates.\cite{Novoselov04,Bonaccorso10,Jablan09,Koppens11,Grigorenko12} Its two dimensional honeycomb lattice structure leads to an effective low-energy theory which is governed by a spin-independent massless Dirac Hamiltonian. The physical implications underlying this fact have been comprehensively studied and spread to a wide extent of different topics.\cite{Neto09} Throughout this article, we are mainly interested in those related to its current current correlation function. In other words, we will continuously allude to the interaction of electrons in graphene with electromagnetic radiation. Some main results covering this aspect have been outlined in Ref. \onlinecite{Peres10}. A constant value for the real part of the conductivity above twice the Fermi energy, i.e., an absorption value equal to 2.3\% dependent only on universal constants for any high enough frequency, deserves to be highlighted.\cite{Mak08,Nair08} At the same time, several issues still demand further research and suggest that this topic is far from being exhausted. For example, the real part of the conductivity shows a plateau between the intra and interband energy regions\cite{Li08} whose numerical value cannot yet be accounted for, theoretically.\cite{StauberPeres08} 

The current response function is an indispensable tool to characterize plasmons. Its tensorial character allows its decomposition in two channels: the longitudinal and transverse channel depending on whether the vector potential $\vec{A}(\vec{q})$ is parallel  or perpendicular to the wave vector $\vec{q}$. Plasmons arising from each channel correspond to Transverse Magnetic or Transverse Electric modes (TM, TE).\cite{Maier07} Those linked to the longitudinal channel have outshined the others, because they are associated to the density response and thus charge accumulation.\cite{Chen12,Fei12} There are also several fundamental studies\cite{Wunsch06,Hwang07} considering the influence of temperature,\cite{Ramezanali09,Stauber12} lattice effects,\cite{Stauber10a,Stauber10c} band gap,\cite{Scholz11} electron-electron interaction\cite{Abedinpour11} or magnetic fields.\cite{Roldan09} Also new plasmon related phenomena have been discussed in double-layer structures, where Coulomb interaction between two graphene layers in the strong light matter coupling regime reveals to be the vehicle that tunnels photons through an otherwise forbidden transmission.\cite{GomezSantos12}

On the other hand, not so much has been clarified about transverse plasmons.\cite{Ziegler07,Jablan11,Pellegrino11,GomezSantos11,StauberPRB12} Ref. \onlinecite{Ziegler07} first spots their existence for the simplest case, namely suspended graphene at zero temperature and making use of the local approximation ($q=0$) for the current current response function. They consist of another kind of oscillation performed by the electrons: whereas for longitudinal plasmons they move forward and backward along the direction determined by $\vec{q}$ (and thus involving charge accumulation), for transverse plasmons the current oscillates in the perpendicular direction and the density is kept homogeneous in space.

Due to their transverse nature, the dispersion relation is closely pinned to the light cone and they are confined to energies between 1.667 and twice the Fermi energy (for larger energies they are strongly damped).\cite{Ziegler07} Here, we want to discuss this transverse light matter coupling in more detail, such as the influence exerted by finite temperature, the possibility of inducing transverse plasmons in the undoped graphene sheet, the circumstances that demand the discard of the local approximation for the response function, the changes induced by modifying the dielectric surroundings, the highest spatial confinement within reach or their behavior when sheets are embedded between an optical cavity.

The paper is organized as follows. In the second section, we calculate the current current response function at finite temperature and doping for both the longitudinal and transverse channels to count on the fundamental tool to discuss light matter interaction within linear response. We then analyze the questions mentioned above: i) effects related to finite temperature. ii) effects related to various dielectric media and the necessity of going beyond the local response. iii) effects related to placing graphene inside a vertical cavity. We complement this discussion by calculating the transmissivity of electromagnetic radiation for $s$ and $p$ polarizations for a single and double layer structure and close with a summary and outlook.

\section{Response functions at finite temperature and doping}

The (bare) response function $\chi_{jj}(\vec{q},\omega)$ ($\chi_{jj}^{(0)}(\vec{q},\omega)$) allows us to determine the value of the currents arising in a medium due to the presence of a total (external) vector potential $\vec{A}(\vec{q},\omega)$ ($\vec{A}^{\text{ext}}(\vec{q},\omega)$). That is, as linear response theory states for an homogeneous system, $\vec{j}(\vec{q},\omega)=\hat{\chi}_{jj}(\vec{q},\omega)\vec{A}(\vec{q},\omega)$ and $\vec{j}(\vec{q},\omega)=\hat{\chi}_{jj}^{(0)}(\vec{q},\omega)\vec{A}^{\text{ext}}(\vec{q},\omega)$, where\cite{Vignale}
\begin{equation}
\hat{\chi}_{jj}(\vec{q},\omega)=
\left(
\begin{array}{c c}
\chi_{j_xj_x}(\vec{q},\omega) & \chi_{j_xj_y}(\vec{q},\omega)\\
\chi_{j_yj_x}(\vec{q},\omega) & \chi_{j_yj_y}(\vec{q},\omega)
\end{array}
\right) \label{matriz}
\end{equation}
(an analogous expression is fulfilled for the bare response, with the superscript $(0)$ in every $\chi$ letter, that is omitted above for the sake of simplicity) and
\begin{equation}
\chi_{AB}^{(0)}(\vec{q},\omega)\equiv -\frac{i}{\hbar S}\lim_{\epsilon\to 0^+}\int_0^\infty dt \langle [\hat{A}_{\vec{q}}(t),\hat{B}_{-\vec{q}}(0)]\rangle e^{i\omega t}e^{-\epsilon t}, \label{Kubo}
\end{equation}
being $S$ the total area of the system.
From now on, every property of the full response function $\hat{\chi}_{jj}(\vec{q},\omega)$ will be shared with the bare one unless the contrary is specified.
$\hat{\chi}_{jj}(\vec{q},\omega)$ is a tensor of the same dimension of the space under study (2 in case of graphene). We can choose a reference frame with one axis parallel to any given $\vec{q}$ in such a way that $\hat{\chi}_{jj}(\vec{q},\omega)$ will become diagonal.\cite{Vignale} In that case, we can decouple the response in two channels: the longitudinal and the transverse one, alluding respectively to $\vec{A}(\vec{q},\omega)$ being parallel or perpendicular to $\vec{q}$. Therefore, $\chi_{j_xj_x}(\vec{q},\omega)$ will provide the result of the longitudinal (transverse) component of the response function when $\vec{q}=q\hat{x}$ ($\vec{q}=q\hat{y}$). Moreover, due to invariance with respect to rotations in our model, these components will only depend on $|\vec{q}|=q$.

To calculate $\chi_{j_xj_x}(\vec{q},\omega)$ for graphene, we will have to use Eq. (\ref{Kubo}) and the Fourier transform of the current operator,
\begin{equation}
\hat{\vec{j}}_{\vec{q}}=ev_F\sum_{\vec{k},\alpha,\beta}\hat{\psi}^\dagger_{\vec{k}-\vec{q},\alpha}\vec{\sigma}_{\alpha\beta}\hat{\psi}_{\vec{k},\beta},
\end{equation}
with $\vec{\sigma}=(\sigma_x,\sigma_y)$ containing the Pauli matrices. It is obvious, then, that
\begin{equation}
\chi_{jj}(\vec{q},\omega)=e^2v_F^2\chi_{\sigma\sigma}(\vec{q},\omega).
\end{equation}
Differing the isospin isospin response function only in a multiplicative factor with respect to the current current one, from now on, we will deal with $\chi_{\sigma_x \sigma_x}(\vec{q},\omega)$. The Lehmann representation yields
\begin{align}
\chi_{\sigma_x \sigma_x}^{(0)}(\vec{q},\omega)=&\frac{1}{S}\lim_{\epsilon\to 0^+}\sum_{\vec{k}}\sum_{\lambda\lambda'}\frac{n_{\vec{k},\lambda}^{(0)}-n_{\vec{k}+\vec{q},\lambda'}^{(0)}}{\hbar\omega+\epsilon_{\vec{k},\lambda}-\epsilon_{\vec{k}+\vec{q},\lambda'}+i\epsilon}\notag\\
&\times |\bra{\chi_\lambda(\vec{k})}\sigma_x\ket{\chi_{\lambda'}(\vec{k}+\vec{q})}|^2. \label{chisigmasigma}
\end{align}
In this expression, we will introduce the wavefunctions
\begin{equation}
\chi_\lambda(\vec{k})=\frac{1}{\sqrt{2}}\left(
\begin{array}{c}
1\\
\lambda e^{i\phi_{\vec{k}}}
\end{array}\right)
\end{equation}
and the non interacting Fermi statistics, $n^{(0)}_{\vec{k},\lambda}$. $\lambda=+1(-1)$ corresponds to the upper (lower) band and $\phi_{\vec{k}}$ is the angle between $\hat{x}$ and $\vec{k}$.

It is also interesting to mention that, by virtue of the continuity equation, the density density and the longitudinal isospin isospin response functions are related, fulfilling \cite{Principi09}
\begin{equation}
\chi_{\rho\rho}(q,\omega)=\frac{v_Fq}{\omega^2}\langle[\sigma^x_{q},\rho_{-q}]\rangle+\frac{v_F^2q^2}{\omega^2}\chi_{\sigma_x\sigma_x}(q\hat{x},\omega). \label{continuidad}
\end{equation} 
On the other hand, the transverse channel is completely independent of the density response.

Counting on all previous considerations, we can write the isospin isospin bare response functions as follows, with $\beta=1$ ($\beta=-1$) for the longitudinal (transverse) channel (that is, we could write $\vec{q}=\frac{q}{2}(|\beta+1|\hat{x}+|\beta-1|\hat{y})$ for $\beta=\{-1,+1\}$ in the left hand side of the formulas below):
\begin{widetext}
\begin{align}
\text{Re} \big(\chi_{\sigma_x\sigma_x}^{(0)}(\vec{q},\omega,T,\mu)\big)&=
-\frac{E_{\text{max}}}{4\pi\hbar^2v_F^2}+(-1)^{\frac{\beta+1}{2}}\frac{g}{4\pi\hbar v_F^2}\sum_{\alpha=\pm}\Bigg[\frac{\omega^2}{q^2}\frac{2k_BT\log\left[1+\exp\left(\frac{\alpha\mu}{k_BT}\right)\right]}{\hbar v_F^2}+ \notag \\
& +\Theta(\omega-v_Fq)f_{\beta}(\omega,v_Fq)\left[G_+^{(\alpha,\beta)}(q,\omega,T,\mu)-G_-^{(\alpha,\beta)}(q,\omega,T,\mu)\right]+ \notag \\
& +\Theta(v_Fq-\omega)f_{\beta}(\omega,v_Fq)\left[\frac{\pi}{2}\Theta(-\alpha)-H^{(\alpha,\beta)}(q,\omega,T,\mu)\right]\Bigg];
\label{reres}
\end{align}
\begin{align}
\text{Im} \big(\chi_{\sigma_x\sigma_x}^{(0)}(\vec{q},\omega,T,\mu)\big)&=
\frac{g}{4\pi\hbar v_F^2}\sum_{\alpha=\pm}\Bigg[ \Theta(v_Fq-\omega)f_{\beta}(\omega,v_Fq)\left[G_+^{(\alpha,\beta)}(q,\omega,T,\mu)-G_-^{(\alpha,\beta)}(q,\omega,T,\mu)\right]+ \notag \\
& +\Theta(\omega-v_Fq)f_{\beta}(\omega,v_Fq)\left[-\frac{\pi}{2}\Theta(-\alpha)+H^{(\alpha,\beta)}(q,\omega,T,\mu)\right]\Bigg]; \label{imres}
\end{align}
\end{widetext}
with
\begin{equation}
f_{\beta}(\omega,v_Fq)=\frac{\omega}{2} \left| 1-\frac{v_F^2q^2}{\omega^2} \right|^{-\beta/2};
\end{equation}
\begin{equation}
G^{(\alpha,\beta)}_\pm(q,\omega,T,\mu)=\int_1^\infty \frac{u\left[ 1-\frac{1}{u^2} \right]^{\beta/2}}{\exp\left[\frac{\hbar|v_Fqu\pm\omega|-2\alpha\mu}{2k_BT}\right]+1}du;
\end{equation}
\begin{equation}
H^{(\alpha,\beta)}(q,\omega,T,\mu)=\int_{-1}^1 \frac{|u|\left[ \frac{1}{u^2}-1 \right]^{\beta/2}}{\exp\left[\frac{\hbar|v_Fqu+\omega|-2\alpha\mu}{2k_BT}\right]+1}du.
\end{equation}

$E_{\text{max}}$ is an ultraviolet cutoff which is canceled by the diamagnetic contribution, as required by Gauge invariance.\cite{Principi09} We include the spin and valley degeneracies as $g=g_sg_v$. Given the electronic density $n$, the chemical potential is determined by
\begin{equation}
\int_{-\infty}^{\infty}d\epsilon\ \nu(\epsilon)\left[n_F(\epsilon)-\Theta(-\epsilon)\right]=n,
\end{equation}
where the density of states is $\nu(\epsilon)=g|\epsilon|/(2\pi \hbar^2v_F^2)$.

These are the main analytical results of this work. They generalize others as the ones given by Refs. \onlinecite{Principi09,Stauber10b}, whose validity is restricted to $T=0$, or those of Ref. \onlinecite{Ramezanali09}, where only the longitudinal channel (or, equivalently, the density density response) is taken into account.

\section{Transverse response and plasmons}

So far, we count on Eqs. (\ref{reres}) and (\ref{imres}) for the bare current current response. Those formulas deal with the case of independent electrons, that is, they cannot manifest the existence of plasmons. However, by means of the Random Phase Approximation (RPA), we can easily obtain a current current response function which takes into account the interaction between electrons. Ref. \onlinecite{Stauber12} assures that without neglecting retardation, we can write
\begin{equation}
\chi_{j_xj_x}^{\text{RPA}}(\vec{q},\omega)=\frac{\chi_{j_xj_x}^{(0)}(\vec{q},\omega)}{1-d^0_{l,t}\chi_{j_xj_x}^{(0)}(\vec{q},\omega)} \label{chiRPA}
\end{equation}
for both the longitudinal ($\vec{q}=q\hat{x}$) and transverse ($\vec{q}=q\hat{y}$) channels knowing the respective propagators $d^0_l$ and $d^0_t$:
\begin{equation}
d_l^0=\frac{q'}{2\epsilon\epsilon_0\omega^2}; \ d^0_t=-\frac{\mu\mu_0}{2q'}.
\end{equation}
In turn,
\begin{equation}
q'=\sqrt{q^2-(n\omega/c)^2}.
\end{equation}

The singularities of Eq. (\ref{chiRPA}) in the longitudinal (transverse) channel will tell when longitudinal (transverse) plasmons arise. Both correspond to collective current oscillations of the electrons, but the character of the oscillations is quite different due to the parallel (perpendicular) direction of their movement with respect to $\vec{q}$ and consequently their (in)depedence on density fluctuations.

Regarding transverse plasmons, the analysis of the denominator of Eq. (\ref{chiRPA}) was carried out by Ref. \onlinecite{Ziegler07}, demonstrating the existence of TE modes in graphene. However, its conclusions are restricted to zero temperature, suspended graphene in vacuum and the local ($q=0$) approximation of the conductivity.

Here, we investigate what happens if those simplifications are discarded. Our standpoint will be able to yield the answer to this and many other questions, as outlined in the introduction. In the following sections, it is our aim to offer a better understanding of TE modes in graphene by means of the result for the current current response functions presented in Eqs. (\ref{reres}) and (\ref{imres}).

\subsection{Transverse plasmons and temperature}

A suitable tool to inquire about the dispersion relation of plasmons is the loss function, which can be defined as
\begin{equation}
S(q,\omega)=-\text{Im}\left(\chi_{j_xj_x}^{\text{RPA}}(\vec{q},\omega)\right). \label{lf}
\end{equation}
Aside from reproducing the singularities of Eq. (\ref{chiRPA}) at $T=0$, it gives information about how they broaden due to damping arising from the bare response (damping coming, in our case, from electron-hole excitations) when $T>0$ or whenever $\text{Im}\left(\chi_{j_xj_x}^{(0)}(\vec{q},\omega)\right)\neq 0$.

It is already known that density plasmons hold (although, of course, damped) for room temperature \cite{Vafek06,Falkovsky07} or higher \cite{Stauber12} ($T\sim T_F$) in the sense that the loss function is not excessively smeared around the singularity present at $T=0$.

\begin{figure*}
\begin{center}
\includegraphics{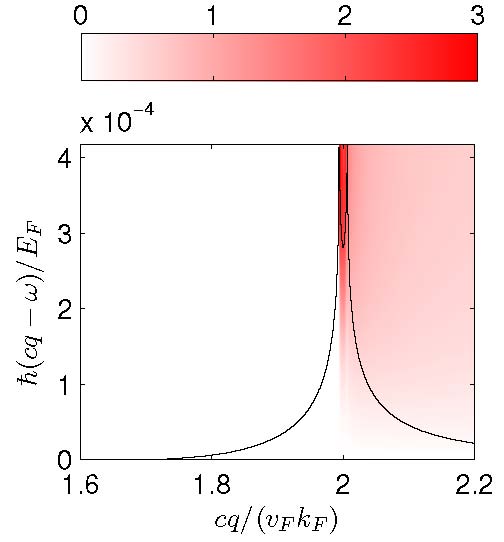} \hspace{1.5cm}
\includegraphics{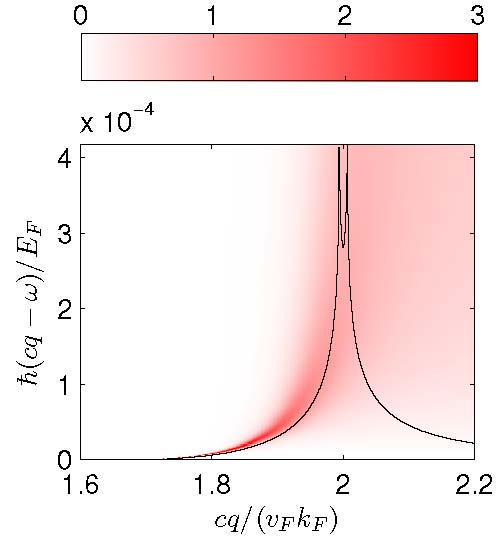}
\end{center}
\caption{Solid lines: dispersion relation for transverse plasmons at $T=0$ (setting $\text{Im}\left(\chi_{j_xj_x}^{(0)}\right)=0$). Color: value of the loss function -Eq. (\ref{lf})- for a given temperature. Left: $T=0$; right: $T=0.02T_F$ (roughly $27K$ for $E_F=0.12\text{eV}$).} \label{lffig1}
\end{figure*}

\begin{figure*}
\begin{center}
\includegraphics{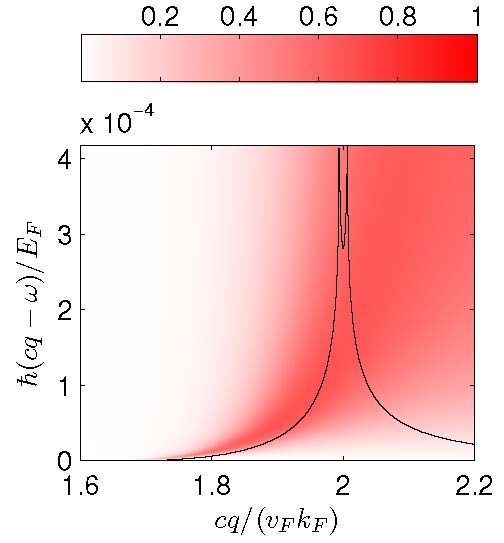} \hspace{1.5cm}
\includegraphics{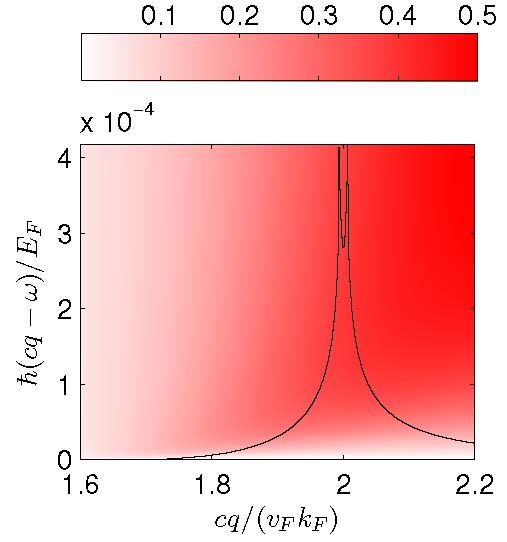}
\end{center}
\caption{Same as in Fig. \ref{lffig1}. Left: $T=0.04T_F$ (roughly $53K$ for $E_F=0.12\text{eV}$); right: $T=0.1T_F$ (roughly $134K$ for $E_F=0.12\text{eV}$). Whereas in the first case the dispersion relation structure is still recognizeable, in the last one it is completely lost.} \label{lffig2}
\end{figure*}

However, transverse plasmons behave in a different manner regarding this aspect. To demonstrate so, let us focus on Figs. \ref{lffig1} and \ref{lffig2}. There, we represent the dispersion relation of transverse plasmons at $T=0$ with a solid line, \footnote{$\text{Im}\left(\chi^{(0)}_{j_xj_x}(q\hat{y},\omega)\right)$ is set equal to 0 for $\omega>E_F/\hbar(2-q/k_F)$, i.e. outside the Pauli blocking zone, where the loss function is really smeared around the solid line plotted because of electron-hole excitations.} whereas the colors correspond to the values of the loss function at the specified temperature. It is convenient to plot the difference of frequency with respect to the light cone $cq-\omega$ in units of $E_F/\hbar$ against the wave vector $q$ in units of $v_Fk_F/c$.

We can see that at very low temperatures, the loss function still suggests the structure of the dispersion relation of transverse plasmons: they are so far well defined. However, rising $T$ to $0.1T_F$ implies that no trace of them is preserved. This is due to the vicinity of the plasmon dispersion to the region of interband  transition and also due to the small spectral weight of the transverse plasmon suppressed by $1/\tilde{c}^2$ with $\tilde{c}=c/v_F\approx300$. The effect of frequency shifting as a consequence of the influence of $T$ is also worthwhile being mentioned.\cite{Stauber12} Not only does temperature determine to which extent longitudinal plasmons are damped, but also displaces the dispersion relation towards higher energies. From Figs. \ref{lffig1} and \ref{lffig2}, an analogous result can be inferred for the transverse channel. Nevertheless, here we observe a red instead of a blue shift; the transverse plasmon thus becomes more localized but finally fades 
 out.

\subsection{Plasmons at zero doping}
Another interesting consideration involving temperature is its ability to induce plasmonic excitations at zero doping. When $E_F$ and $T$ equal 0, neither electric nor magnetic modes can be present in a graphene layer. However, finite temperature involves thermally activated electron-hole excitations, allowing longitudinal (slightly damped) plasmons to appear.\cite{Vafek06} In fact, the case of $E_F=0$ and $T>0$ ($T$ being sufficiently low) can be shown to be equivalent to doped graphene at $T=0$ with the Fermi energy $E_T\equiv 2\ln2k_BT$.\cite{Falkovsky07} In other words, the role of temperature is equivalent to inducing a nonzero value of doping.

Thus basically two forces compete in the context of this mechanism to make plasmons spring up: the excitation of carriers being favorable to their emergence and the increase of damping claiming to make them vanish. Whereas in the longitudinal case there are well-defined oscillations, we find that for transverse plasmons they are completely washed out. We can understand this by comparing the energy scale set by the temperature $T$ with the energy scale given by $E_T$:
\begin{equation}
k_BT/E_T=1/(2\ln2)\simeq 0.7 \label{scale}\;.
\end{equation}
Since already for $T=0.1T_F$ there is no clear maximum in the loss function (see Fig. \ref{lffig2}), the scale set by Eq. (\ref{scale}) seems too high to induce transverse plasmons at zero doping.

\begin{figure}
\begin{center}
\includegraphics{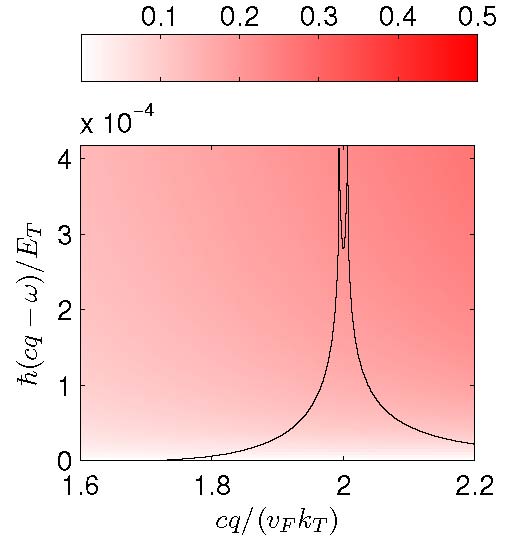}
\end{center}
\caption{Same as in Fig. \ref{lffig1}, but for $E_F=0$ and $T=1K$. The solid line corresponds to $k_T=E_T/(\hbar v_F)$ and $T=0$, with $E_T\equiv 2\ln2k_BT$. The loss function does not lead to any trace of transverse plasmons.}  \label{lfEF0}
\end{figure}

Fig. \ref{lfEF0} confirms this intuition: the loss function for $T=1K$ is completely diluted and does not reproduce the plasmon dispersion relation respective to $T=0$ and a nonzero doping given by $E_T$. Lowering the temperature even more does not involve any change, which can be expected since Eq. (\ref{scale}) is a scale-invariant universal result. Thus, we have reached a remarkable conclusion concerning a difference between transverse and longitudinal plasmons: only the latter can be found at zero doping.

\section{Influence of dielectric media}

It is well known how longitudinal plasmons change their dispersion relation when graphene sheets (e.g. monolayer and double layer systems) are embedded between different dielectrics.\cite{Stauber12} However, transverse plasmons do not behave similarly even in the simplest case, that is, a single sheet lying on a substrate. They exhibit an extreme sensibility to a slight difference in the refractive index of the two surrounding media to the extent of vanishing for $|n_2-n_1|\sim 10^{-7}$ at room temperature.\cite{Kotov13} This is due to the fact that the dispersion relation of transverse plasmons is extremely pinned to the light cone, such that when two different light cones exist and are sufficiently separated, they rapidly vanish.

Thus to focus on transverse plasmons and inquire about how the dispersion relation can substantially change due to their dielectric surroundings (permittivity and permeability), we will keep the vicinity of the graphene layer homogeneous with the same refractive index $n$. Within this constraint, we can analyze the consequences of modifying $n$.

\begin{figure}
\begin{center}
\includegraphics{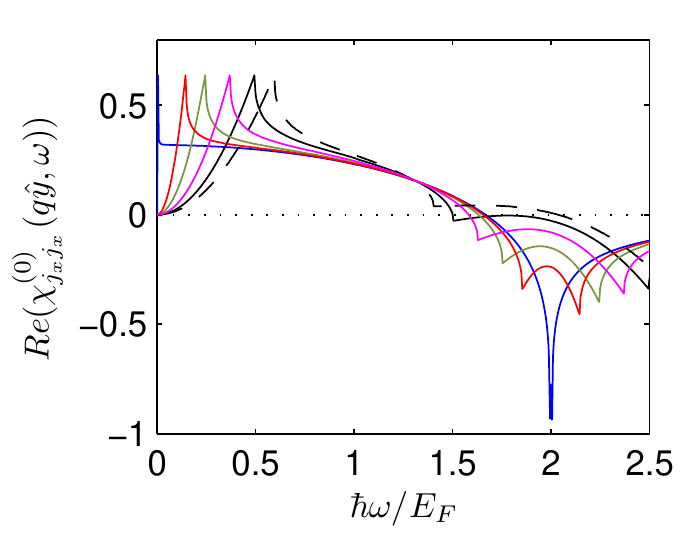}
\end{center}
\caption{$\text{Re}\left(\chi_{j_xj_x}^{(0)}(q\hat{y},\omega)\right)$ at $T=0$ for different values of $q/k_F$: $5\cdot 10^{-3}$ (blue), $0.15$ (red), $0.25$ (green), $0.37$ (magenta), $0.5$ (solid black) and $0.6$ (dashed).} \label{rechijjqyqgrande}
\end{figure}

\subsection{Influence of the refractive index}
Increasing the refractive index $n$ decreases the speed of light such and as a consequence the light cone is shifted to greater values of $q$ for a fixed energy $\hbar\omega$. Since the plasmon relation must lie in the evanescent region where $\omega<cq/n$, the current current response $\chi_{j_xj_x}^{(0)}(q\hat{y},\omega)$ involved in the present situation must be evaluated at larger $q$. As we mentioned before, previous discussions of the dispersion relation as the one in Ref. \onlinecite{Ziegler07} would not be valid for high enough values of $n$, since they are only based on the local response ($q=0$). The relevance of this dependence is shown in Fig. \ref{rechijjqyqgrande}, where the real part of the transverse current response is shown as a function of the energy for $q/k_F\sim 0-0.6$.  The remarkable differences between curves imply significant changes in $\omega(q)$. Thus Eqs. (\ref{reres}) and (\ref{imres}) will provide us with the ability of characterizing transverse plasmons for high values of the refractive index.

\begin{figure*}
\begin{center}
\includegraphics{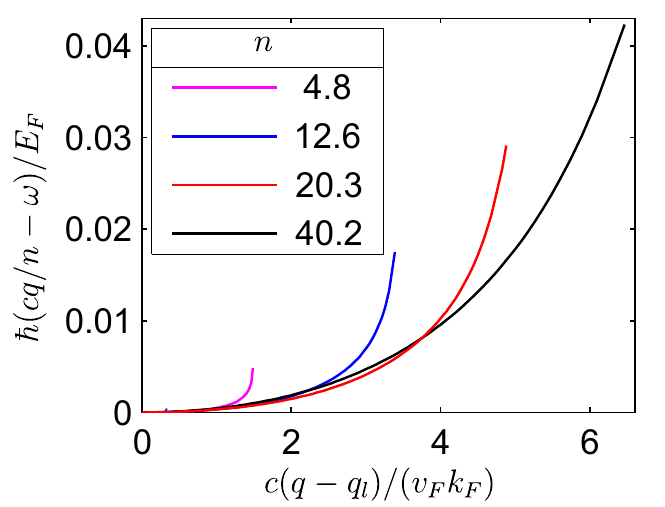} \hspace{1.5cm}
\includegraphics{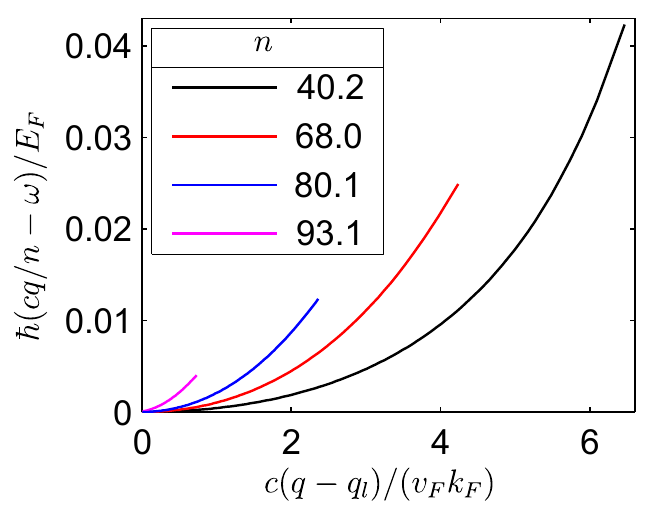}
\end{center}
\caption{Dispersion relation of transverse plasmons for different refractive index $n$ of the homogeneous dielectric medium embedding graphene. We only show the undamped region at $T=0$. Curves are normalized to $q_{l}=n\omega_{l}/c$ with $\omega_{l}$ the lowest frequency for which transverse plasmons appear.} \label{rdqgrande}
\end{figure*}

Some results are shown in Fig. \ref{rdqgrande}, where we focus on undamped plasmons at $T=0$. They can be compared to the solid black line of Figs. \ref{lffig1}-\ref{lfEF0}, respective to $\epsilon\mu=1$. It is interesting to notice that, as we increase $n$ starting from $n=1$, the highest value of the curves moves further and further away from the light cone (left hand side of Fig. \ref{rdqgrande}). But if we keep up increasing $n$, it turns back (right hand side of Fig. \ref{rdqgrande}). Therefore, there is a maximal separation of the dispersion relation from the light cone which determines the fastest decay possible of the fields away from the graphene sheet. \cite{Stauber12} The confinement of transverse plasmons can thus be increased by a factor of $10^2$ by means of embedding the layer within a dielectric with $n\simeq 40$.

\begin{figure}
\begin{center}
\includegraphics{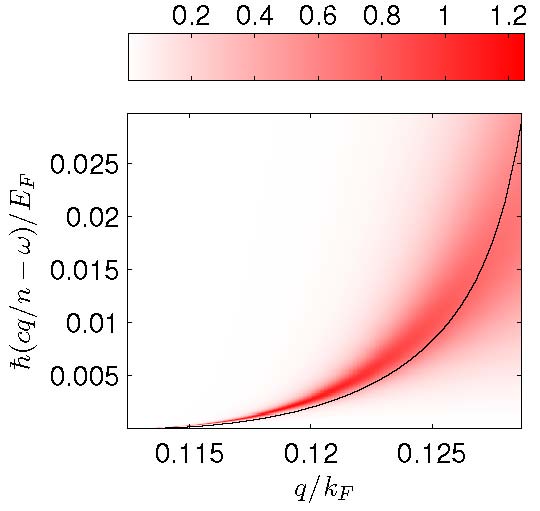}
\end{center}
\caption{Same as Fig. \ref{lffig1}, but for $T=0.02T_F$ and $n=20.3$. The loss function indicates the presence of damped transverse plasmons.} \label{lfqgrande}
\end{figure}

Regarding now the influence of temperature, there are no substantial differences with respect to the case $\epsilon\mu=1$ (previous section). Fig. \ref{lfqgrande} represents the loss function, which allows to see how at low temperatures the structure of the dispersion relation of plasmons is maintained. Once again, for $T>0.1T_F$, it is completely diluted. Following the reasoning stated in Eq. (\ref{scale}), neither will it be possible to have transverse plasmons for $E_F=0$ induced by a finite $T$.

\subsection{Influence of an optical cavity}

Density plasmons are an excellent way of enclosing radiation in small regions. These dimensions just depend on the separation of the dispersion relation from the light cone: the decay length can be written as $\lambda=2\pi/q'$, with $q'=\sqrt{q^2-(n\omega/c)^2}$.\cite{Stauber12} For energies of the order of $E_F/\hbar$, we can neglect retardation effects ($q'\simeq q$), yielding $\lambda\sim 2\pi/k_F$ and decay lengths of the order of $10\text{nm}$.

However, once again the situation turns inside out when considering the transverse channel. The proximity to the light cone implies that $q'$ (even its maximum value, which can be estimated from Fig. \ref{rdqgrande}) is now much smaller than that of its longitudinal counterpart. For suspended graphene, for example, we find a minimum value for undamped plasmons of $\lambda\sim 10^{-4}\text{m}$. It is our aim to address the question of to which extent dodging this limitation is possible and whether confinement can be achieved by relying on a multilayer system.

\begin{figure}
\begin{center}
\includegraphics{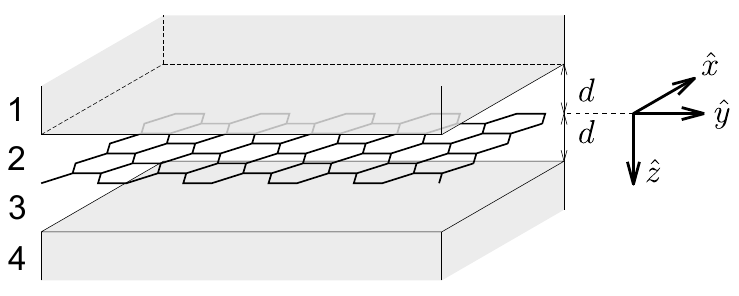}
\end{center}
\caption{Schematic view of the cavity with graphene suspended in air. Media 1 and 4 are semi infinite dielectrics/superconductors.} \label{esquema}
\end{figure}

Let us thus consider a single sheet of graphene embedded between four dielectrics (Fig. \ref{esquema}). For the sake of simplicity, we will set $\mu_1=\mu_4\to 0$  and $\mu_2=\mu_3=1$, but keep the velocity of light constant and equal within the whole sample. This makes media 1 and 4 impenetrable by $s$ polarized electromagnetic waves and one might expect to force a faster decay of the potential vector $\vec{A}$ in $z$-direction.

To analyze this problem, we make the ansatz
\begin{align}
\vec{A}_{j}(\vec{r},t)&= M_j\hat{y} \exp \left[-q' z+i\left(qx-\omega t\right)\right]+\notag \\
&N_j\hat{y} \exp \left[q' z+i\left(qx-\omega t\right)\right] \label{quinit}
\end{align}
for every medium $j$, setting $M_1=N_1=M_4=N_4=0$, and apply Maxwell equations in every interface. The transverse current current response of graphene arises when attending to the frontier between dielectrics 2 and 3 (vacuum in our case), as well as RPA turns up when writing the surface current as\cite{Vignale} $\vec{j}(\vec{q},\omega)=\hat{\chi}_{jj}^{(0)}(\vec{q},\omega)\vec{A}(\vec{q},\omega)$ (notice the presence of $\hat{\chi}_{jj}^{(0)}$ instead of $\hat{\chi}_{jj}$). The resulting system of linear equations for $\{M_j,N_j\}$ with $j=\{2,3\}$ will have a nontrivial solution only when its determinant is zero ($d>0$):
\begin{equation}
2q'+\mu_0\chi_{j_xj_x}^{(0)}(q\hat{y},\omega)\tanh(dq')=0. \label{det}
\end{equation}
In that case,
\begin{equation}
N_2=M_3=-e^{2dq'}M_2; \ N_3=M_2;
\end{equation}
where $M_2$ is related to the field amplitude.

Let us now define the scale for the decay length. Denoting $q_p'$ as the retarded wave vector which is most separated from the light cone and thus related to the maxima in Fig. \ref{rdqgrande}, we can define $\lambda=2\pi/q'_p$ as the length scale within the transverse plasmon will be confined. Solving Eq. (\ref{det}) numerically, we find a solution only for layer separation $d>0.15\lambda$. Thus, transverse plasmons in suspended graphene can be maximally confined to length scales of the order of $10^{-5}\text{m}$, still much larger than the previously commented associated to their longitudinal homologous. Relaxing the boundary conditions to other values does not significantly change these conclusions.

\begin{figure}
\begin{center}
\includegraphics{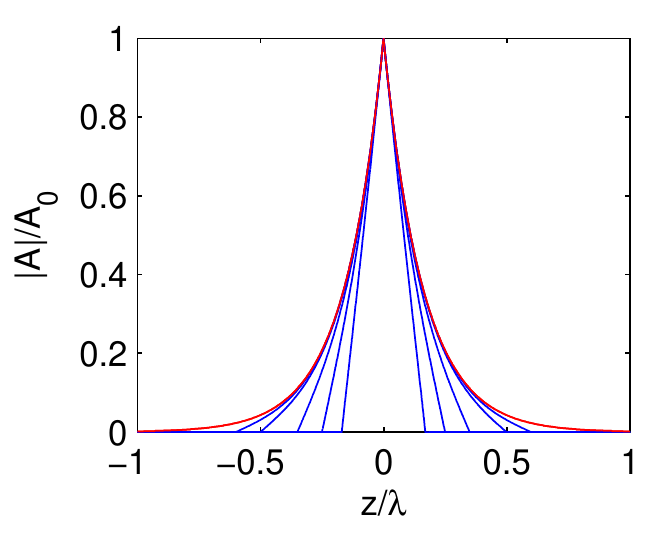}
\end{center}
\caption{Modulus of confined vector field of transverse plasmons for graphene in a cavity. Blue curves correspond to $d/\lambda$ equal to 0.17, 0.25, 0.35, 0.5 and 0.6 (starting from the inner curve). For $d\leq 0.15\lambda$, no confined modes exist. Red curve: graphene in free space (media 1 and 4 absent). The length scale is given by $\lambda=2.65\cdot 10^{-4}\text{m}$.} \label{decays}
\end{figure}

For short enough distances, it can be seen that the decay in the perpendicular direction to the frontiers differs quite a lot from being exponential, rather becoming practically linear. The plot for several values of $d$ appears in Fig. \ref{decays}, closing our discussion about transverse plasmons regarding the spatial decay of the electromagnetic fields attached to them.

To sum it up, this section has mainly remarked the disparity between longitudinal and transverse plasmons as for confinement in the outskirts of the sheet, being the latter significantly more spread in space. This emerges from the nature underlying their dispersion relation (namely its proximity to the light cone) and can be hardly eluded even with the aid of setups like the one with impenetrable dielectrics shown in Fig. \ref{esquema}.

\section{Absorption in single and double layer systems}

In the previous sections, we have highlighted some aspects concerning the evanescent spectrum. Here, we will also cover some others related to the propagating modes, i.e., the absorption of $s$ and $p$ polarized light by a single and double layer graphene structure. 

\begin{figure*}
\begin{center}
\includegraphics{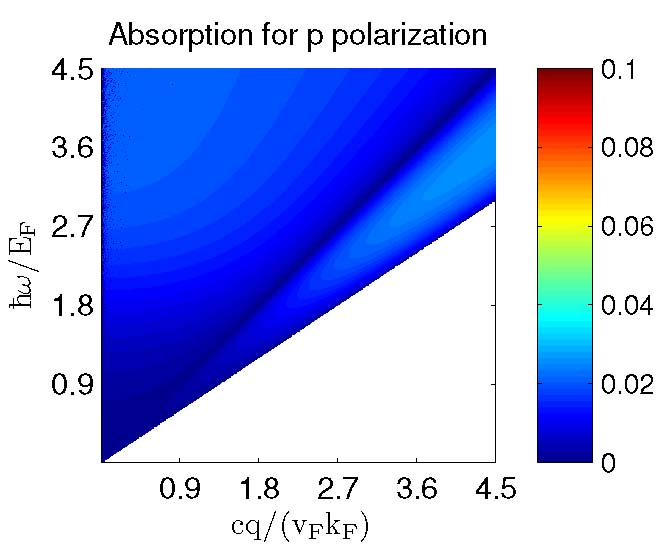} \hspace{0.5cm}
\includegraphics{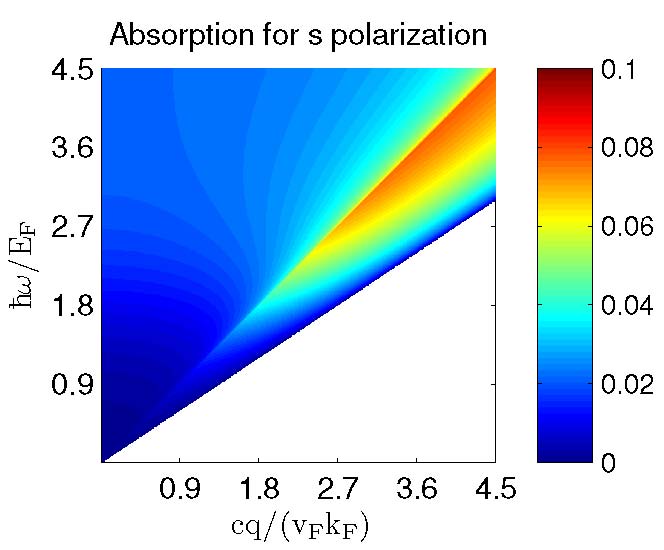}
\end{center}
\caption{Absorption for graphene on a substrate with $\mu=1$ and $n=1.5$ for $p$ and $s$ polarized light at $T=300K$. Incidence occurs from the substrate such that there is a critical angle for total reflection.} \label{abs}
\end{figure*}

\subsection{Single layer structures}

The amount of energy transmitted, reflected and absorbed is encoded in the Fresnel coefficients which emerge from the application of Maxwell equations in the frontier between two dielectrics when a layer of graphene is placed separating them.\cite{FalkovskyPer07,Gaskell09} For $s$ polarized light, the Fresnel coefficients for the parallel (conserved) component are given by\cite{Stauber12}
\begin{align}
t_s&=\frac{2\mu_2q_1^\prime}{\mu_2q_1^\prime+\mu_1q_2^\prime+\mu_1\mu_2\mu_0\chi_{j_xj_x}^{(0)}(q\hat{y},\omega)}\;,\\
r_s&=\frac{\mu_2q_1^\prime-\mu_1q_2^\prime-\mu_1\mu_2\mu_0\chi_{j_xj_x}^{(0)}(q\hat{y},\omega)}{\mu_2q_1^\prime+\mu_1q_2^\prime+\mu_1\mu_2\mu_0\chi_{j_xj_x}^{(0)}(q\hat{y},\omega)}\;.
\end{align}
The absorption is then:
\begin{equation}
\mathcal{A}_s=1-|r_s|^2-|t_s|^2\frac{\mu_1q_2'}{\mu_2q_1'}\;.
\end{equation}
Equivalent equations are obtained for $p$ polarized light by replacing $\mu_i\to q_i'$, $q_i'\to\epsilon_i$ and $\chi_{j_xj_x}^{(0)}(q\hat{y},\omega)\to\chi_{j_xj_x}^{(0)}(q\hat{x},\omega)$.\footnote{Note that these expressions slightly differ from standard textbook notation.}

Their application to the simplest case, i.e., suspended graphene, reveals that the maximum absorption ($2.3\%$) for $p$ polarization corresponds to the normal incidence, whereas when dealing with $s$ polarization, it is reached when
\begin{equation}
\left(\frac{qc}{\omega}\right)^2=1-\left(\frac{\pi\alpha}{2}\right)^2; \ \alpha=\frac{e^2}{4\pi\epsilon_0\hbar c}.
\end{equation}
The angle of incidence is given by $\sin \theta=qc/\omega\Rightarrow\theta\simeq 89º$, and the absorption is exactly $50\%$ which resembles the theoretically highest absorption by a single interface.\cite{Thongrattanasiri12}

Also for graphene on a substrate, differences between both polarizations are important close and beyond the light cone. Fig. \ref{abs} shows the results for a dielectric with $\mu=1$ and $n=1.5$. It can be realized that in the zone where total reflection should take place in the absence of graphene, some absorption ($\simeq 3\%$) is found for $p$ polarization, but more than twice as much for $s$ polarization.

\subsection{Double layer structures}

\begin{figure}
\begin{center}
\includegraphics{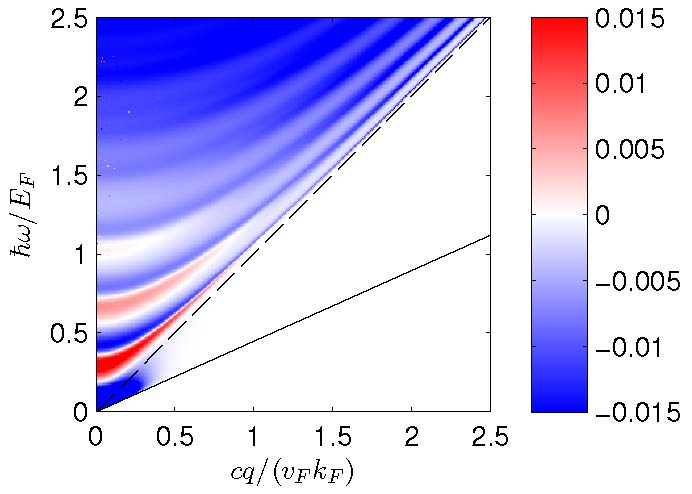}
\end{center}
\caption{Transmission of incident light through a double layer graphene structure with dielectric media $\epsilon_1=\epsilon_3=5$, $\epsilon_2=1$ and $\mu=1$. Shown is the difference $T_g-T_0$, with $T_g$ ($T_0$) the transmission with (without) graphene. The distance separating graphene layers is $d=14.8\mu\text{m}$. The light cones respective to media 1 and 2 are denoted by the solid and dotted black lines, respectively.} \label{noperf}
\end{figure}

Other setups with more layers may also yield interesting results. In Ref. \onlinecite{Stauber12}, two sheets of graphene separating three dielectrics with permittivities $\epsilon_1$, $\epsilon_2$ and $\epsilon_1$ ($\epsilon_1>\epsilon_2$) give rise to perfect transmission between the two light cones in the regime of strong light-matter coupling (for energies of the order of the fine-structure constant times the Fermi energy). The results can be interpreted as surface plasmon mediated extraordinary transmission similar to the one through sub-wavelength apertures.\cite{Ebbesen98} This analysis was carried out for $p$ polarization and here we want to extend it to $s$ polarization. Results are shown in Fig. \ref{noperf}, where no perfect transmission shows up.\footnote{The colored region inside the light cone for $q\to 0$ is due to high reflectivity of graphene at low frequencies.} 

Our interpretation is the aforementioned fragility of transverse plasmons, which vanish for large layer separation in this staging unless the refractive index $n_1$ and $n_2$ are sufficiently close to each other. As a consequence, there is no mean of tunneling $s$ polarized light through transverse plasmons coupling between layers.

\section{Conclusions}

In summary, we have calculated the linear current current response function for graphene at finite temperature and chemical potential. This analytical result enables the characterization and study of plasmons from a quite general standpoint.

This paper was focused on the transverse channel, whose collective oscillations remained not as well studied as the ones of the longitudinal channel. As for them, we have analyzed the strong influence that temperature exerts when compared to their counterparts, in the sense that they vanish much earlier when increasing $T$. Red shifting of the dispersion relation or the unfeasibility of inducing their existence at zero doping due to a finite temperature also make up also clear differences.

Moreover, the influence of the dielectric surroundings on transverse plasmons has been targeted. The main consequences are enclosed in the evolution of the dispersion relation as a result of modifying the refractive index in which graphene is embedded. We have described these curves for a wide range of values of the refractive index which made it necessary to go beyond local response and showed that maximal confinement is obtained for $n\simeq 40$. This confinement can be enhanced by placing graphene inside a cavity consisting of a perfect diamagnet, i.e., $\mu=0$. It reaches a maximum for a certain distance below which no transverse plasmons can be sustained which can be used for a sensor.
 
In the last section, we commented on some aspects of the absorption of electromagnetic radiation by a single and double layer system due to the presence of graphene, where the polarization of light as well as the incident angle give rise to fluctuations up to $\sim50$\%.

\section{Acknowledgments} We thank Guillermo G\'omez-Santos, Reza Asgari and Rafael Rold\'an for helpful discussions. This work has been supported by FCT under grants PTDC/FIS/101434/2008; PTDC/FIS/113199/2009 and MIC under grant FIS2010-21883-C02-02. Ángel Gutiérrez acknowledges financial support through grants JAE-Pre (CSIC, Spain) and accommodation in Residencia de Estudiantes through Ayuntamiento de Madrid.

\end{document}